\documentclass{article}
\usepackage{ijcai24}

\usepackage{times}
\usepackage{soul}
\usepackage{color}
\usepackage{url}
\usepackage[hidelinks]{hyperref}
\usepackage[utf8]{inputenc}
\usepackage[small]{caption}
\usepackage{graphicx}
\usepackage{amsmath}
\usepackage{amsthm}
\usepackage{booktabs}
\usepackage{algorithm}
\usepackage{algorithmic}
\usepackage[switch]{lineno}
\usepackage{xcolor}
\usepackage{colortbl}

\usepackage{newfloat}
\usepackage{listings}
\DeclareCaptionStyle{ruled}{labelfont=normalfont,labelsep=colon,strut=off} 
\floatstyle{ruled}
\newfloat{listing}{tb}{lst}{}
\floatname{listing}{Listing}

\title{A-JEPA: Joint-Embedding Predictive Architecture Can Listen}

\author{
    Zhengcong Fei, Mingyuan Fan, Junshi Huang\thanks{The corresponding author.}
    \affiliations
    Kunlun Inc.
    \emails
    feizhengcong@gmail.com
}

\begin{document}

\maketitle

\begin{abstract}
    This paper presents that the masked-modeling principle driving the success of large foundational vision models can be effectively applied to audio by making predictions in a latent space. We introduce Audio-based Joint-Embedding Predictive Architecture (A-JEPA), a simple extension method for self-supervised learning from the audio spectrum. Following the design of I-JEPA, our A-JEPA encodes visible audio spectrogram patches with a curriculum masking strategy via context encoder, and predicts the representations of regions sampled at well-designed locations. The target representations of those regions are extracted by the exponential moving average of context encoder, \emph{i.e.}, target encoder, on the whole spectrogram. We find it beneficial to transfer random block masking into time-frequency aware masking in a curriculum manner, considering the complexity of highly correlated in local time and frequency in audio spectrograms. To enhance contextual semantic understanding and robustness, we fine-tune the encoder with a regularized masking on target datasets, instead of input dropping or zero. Empirically, when built with Vision Transformers structure, we find A-JEPA to be highly scalable and sets new state-of-the-art performance on multiple audio and speech classification tasks, outperforming other recent models that use externally supervised pre-training. 
\end{abstract}

\section{Introduction}

\begin{figure}[t]
   \centering
   \includegraphics[width=1\linewidth]{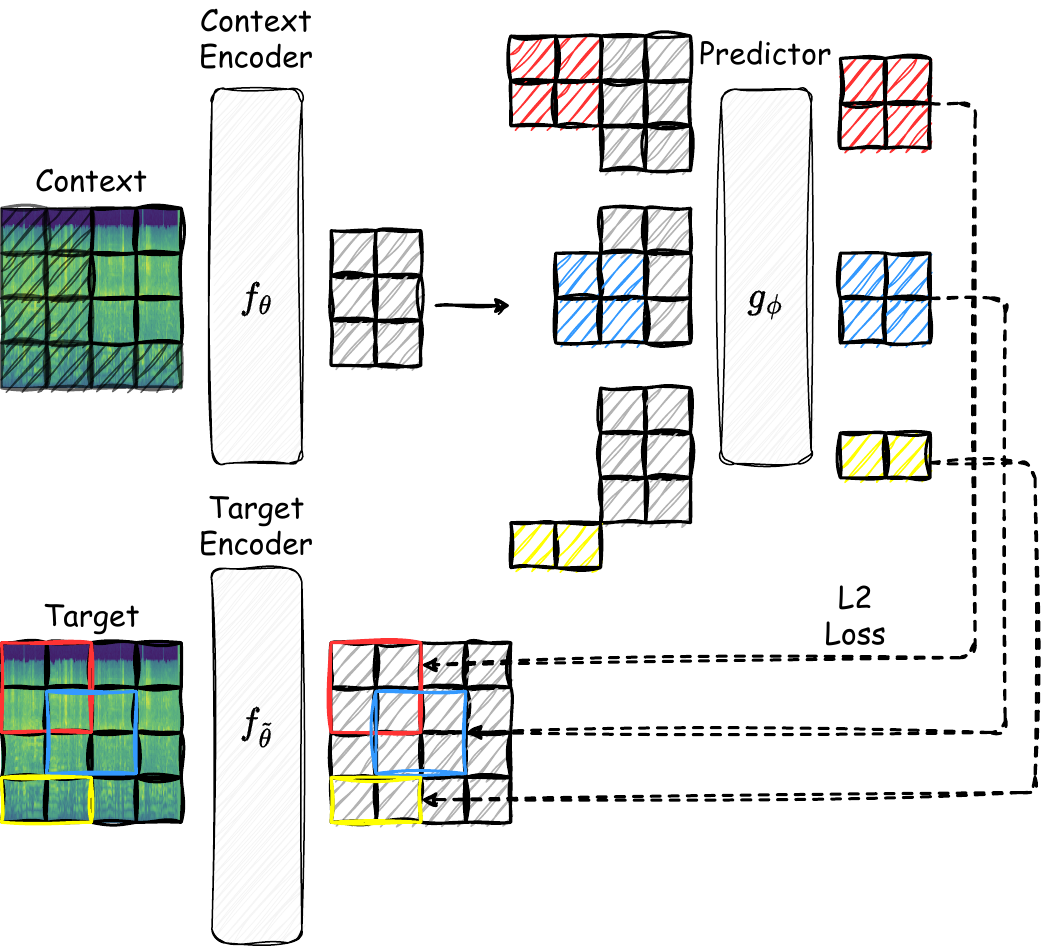}
   \caption{\textbf{Overview of A-JEPA.} The audio-based joint-embedding predictive architecture uses a context encoder to predict and align the representations of various target blocks in a latent space, originating from the same audio spectrogram. 
   }
   \label{fig:1}
\end{figure}

Numerous cognitive theories state that the process of adapting an internal model to predict the absence of input information is a crucial mechanism for learning in biological systems \cite{rao1999predictive}.
The recent accomplishments of substantial foundational language models have been propelled by the utilization of the self-supervised mask-denoising paradigm, which involves learning through the process of filling in missing information \cite{devlin2018bert,brown2020language,touvron2023llama}. 
Moreover, masked pretraining tasks also dominate the performance of representation learning in computer vision \cite{bao2021beit,he2022masked,zhou2021ibot,fei2023masked}.

The recently proposed image-based Joint-Embedding Predictive Architecture (JEPA) \cite{lecun2022path,assran2023self} demonstrates that one could learn well-performed image representations by predicting the masked image regions in a latent space.
Compared to other methods for masked image modeling \cite{he2022masked,fei2019fast,tong2022videomae,fei2022deecap,fei2020iterative,yan2021semi,feichtenhofer2022masked}, which predict low-level visual tokens or pixels, JEPAs make predictions in a high-level representation space, where unnecessary pixel-level details can be eliminated, thereby leading the model to concentrate on more semantic features \cite{assran2023self}. However, the implementation of this learning principle on sensory data, \emph{e.g.}, audio, continues to be a promising endeavor.

In this work, we study the problem of self-supervised representation learning on audio data and extend the JEPA-based learning principle to the audio spectrogram, which we refer to as A-JEPA: Audio-based Joint-Embedding Predictive Architecture. 
Different from JEPA for images, it is believed that the random block masking of audio spectrogram may be easy due to the correlation of information along the time and frequency axis \cite{tong2022videomae}. Therefore, the region masking strategy of A-JEPA is designed in a curriculum manner for the spectrogram of audio signal, \emph{i.e.}, gradually from random block to time-frequency aware masking in a schedule. 
After that, we train the context-encoder by restoring the missing regions of masked spectrogram with the guidance from exponential moving averaged target-encoder in learned representation space. 
We minimize the patch-normalized mean square error to optimize the networks. 
At the fine-tuning stage, we discard the decoder and fine-tune the context encoder with regularized patch-masking, where the attention weights for the masked token are solely reliant on others rather than directly dropped patches or set to zero. 
We target to manipulate the connections between audio patches via masking, where the networks are forced to exploit partial neighbors’ information to produce a meaningful representation.

Empirically, the A-JEPA model exhibits exceptional performance in multiple audio and speech classification tasks, thereby establishing a new state-of-the-art benchmark. 
Notably, as a self-supervised audio-only model, it surpasses other contemporary competitors focused on masking and reconstruction by a significant margin of +1.3 points in terms of mAP on the AudioSet-2M dataset. We further provide the visualization and audible examples to qualitatively ascertain the efficacy of the A-JEPA. 
More encouragingly, our findings indicate that increasing the pre-training audio dataset size yields continued performance enhancements, even under fixed computational constraints. This observation suggests a promising avenue for further advancements in audio foundation models. 
Through an extensive empirical evaluation, we demonstrate that:
\begin{itemize}
    \item This paper delves into a simple extension of JEPA to audio data, presenting a unified and scalable framework for learning self-supervised audio representations. 
    To the best of our knowledge, it is the initial endeavor to apply JEPA to the audio area, yielding remarkably favorable outcomes. 
    \item To cover the unique challenges of the audio domain, we introduce a curriculum masking strategy for gradually time-frequency aware pattern during pre-training, and regularized patch masking for robust information flow during fine-tuning. 
    \item  Experimental results substantiate the scalability and efficiency of the A-JEPA framework. Moreover, A-JEPA exhibits superior performance compared to pixel-reconstruction techniques like AudioMAE in AS-2M classification and other downstream tasks. Finally, the code and models will be publicly available. 
\end{itemize}

\section{Related Works}

\paragraph{Joint-embedding architectures for masked pre-training.}
Masked and denoising autoencoders~\cite{dae,sae,bert} have emerged as versatile methods for learning representations by reconstructing the original source from masked or corrupted inputs \cite{feiincorporating,fei2023uncertainty}.
In CV, a set of approaches attempt to integrate joint-embedding architectures with reconstruction-based approaches, wherein they employ an invariance pretraining loss with a patch-level reconstruction loss \cite{zhou2021ibot,PathakKDDE16,chen2020,mae,maskedfeat}. 
However, it is noteworthy that view-invariance-based methods often exhibit a bias towards learning global image representations. Adding local loss terms is proposed to improve performance on other popular tasks in computer vision \cite{bardes2022vicregl,chen2022intra,gidaris2020learning}. 
The concept of contrastive predictive coding \cite{oord2018representation} is also closely connected to this research direction on local loss terms. In the context of images \cite{henaff2020data}, it uses a contrastive objective combined with a convolutional network to discriminate between overlapping image patch representations. 
Furthermore,  I-JEPA \cite{assran2023self} aims to predict the representations of different target blocks from a single context block, thereby constructing semantic information.  
Our work extends the JEPA framework with unique designs for representation learning with audio spectrograms.

\paragraph{Out-of-domain and in-domain pre-training for audio.} 
Pre-training for audio representation can generally be divided into two main categories: 
(\textbf{i}) Transferring natural image supervised pre-trained ViT~\cite{dosovitskiy2020image} or ResNet~\cite{He2015DeepRL}, \emph{e.g.}, ImageNet ~\cite{gong2021ast,paast,Nagrani21c,chen2022hts,gong2021psla,cmkd}. 
In this approach, the models operate over audio spectrograms by deflating from three channels (RGB) into one channel (spectrogram) in the pre-trained patch embedding in ViT and employing the rest of the transformer blocks \cite{gong2021ast,paast}. 
\cite{chen2022hts} encodes spectrograms initialized from the Swin Transformer~\cite{swin} and 
\cite{Nagrani21c} uses ImageNet-21K pre-trained ViT. 
Instead, our A-JEPA focuses on audio-only self-supervised pre-training from scratch. 
(\textbf{ii}) Audio-only self-supervised methods, which can be further split by the input signal type, \emph{e.g.}, raw waveform~\cite{wav2vec,wav2vec2,d2v}, frame-level features~\cite{Hsu2021HuBERTSS,avhubert,srivastava2021conformer}, or spectrogram patches~\cite{ssast,baade2022}); and the objective used for self-supervision, \emph{e.g.}, contrastive~\cite{cpc,wav2vec2,object_sound,stica,Hsu2021HuBERTSS} or prediction and reconstruction~\cite{ssast,d2v,srivastava2021conformer,avhubert}.
\cite{wav2vec2} takes raw waveform as inputs and exploits contrastive learning to differentiate contextualized representations across different time segments.
\cite{mockingjay} introduces a pretext task of masked acoustic model, aiming to reconstruct frame-level Mel-features from masked time frames. 
\cite{ssast,baade2022,chong2022masked,niizumi2022masked,huang2022masked} operates over spectrogram patches and employs joint contrastive and reconstructive objectives on masked patches. 
Compared with prediction at a low level, we build in a latent space and showcase its superiority.

\section{Approach}

In this section, we review the standard joint-embedding predictive architecture (JEPA) and describe how we instantiate it for the audio domain as A-JEPA.

\subsection{Joint-Embedding Predictive Architecture} 
The main idea of JEPA is to learn an encoder by predicting parts of input, \emph{i.e}, target regions, based on the visible context in the latent space. 
The basic architecture is made up of three networks: a context encoder $E_\theta(\cdot)$ and target encoder $E_{\tilde{\theta}}(\cdot)$ extract the representations of context and target regions, respectively. 
The predictor $P_\phi(\cdot, \cdot)$ is used to predict the target representation condition on the context representation from $E_\theta(\cdot)$.
The parameters of target encoder $E_{\tilde{\theta}}(\cdot)$ are updated via the exponential moving average of context encoder weights at each iteration.
The parameters of context encoder and predictor are optimized jointly by minimizing the distance of representations of target regions from predictor and target encoder. 

\subsection{Application to Audio}

We instantiate the A-JEPA for audio by using a curriculum masking strategy in pre-training and regularized masking in fine-tuning. 
Generally, the context encoder extracts the representation of visible audio spectrogram pre-processed by our masking strategy, based on which the predictor outputs the representation of target regions in latent space. 
See Figure ~\ref{fig:1} for an overview. 

\paragraph{Input spectrogram.}

In line with \cite{gong2021ast,gong2022ssast,huang2022masked}, we transform audio signals into Mel-spectrograms and segment them into non-overlapped grid patches. These patches are then flattened and embedded by a linear projection. We also add fixed sinusoidal positional encoding to the patch embedding.

\begin{figure}[t]
  \centering
   \includegraphics[width=1\linewidth]{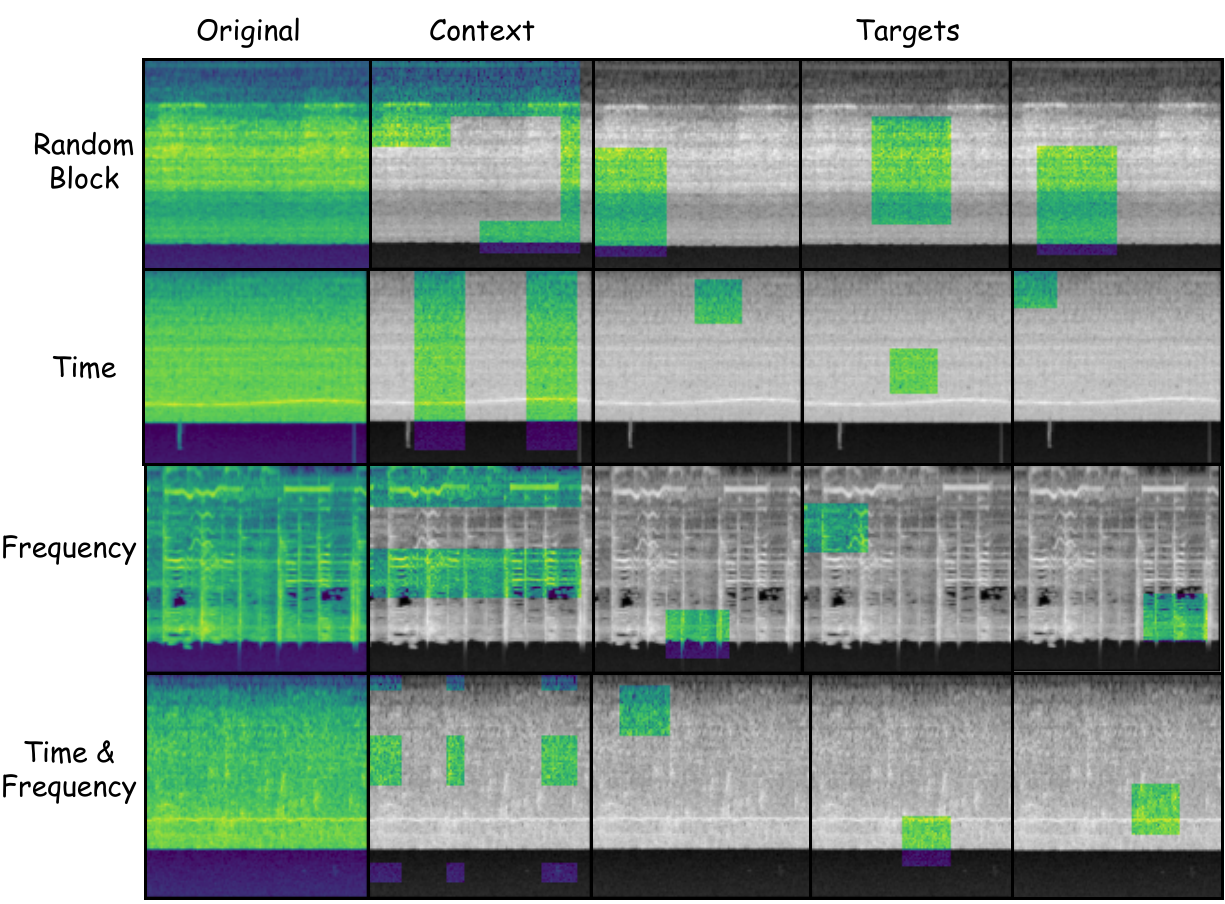}
   \caption{\textbf{Examples of our context and target-masking strategy on Mel-spectrograms.}  Given an audio spectrogram, (i) random block: we randomly sample 4 target blocks with scale in the range (0.15, 0.2) and aspect ratio in the range (0.75, 1.5); (ii) time and frequency: we randomly sample 3 target blocks with scale in the range (0.05, 0.075) and \emph{remove all the related time or frequency} in the total Mel-spectrograms. Next, we randomly sample a context block with a scale in the range (0.85, 1.0) and \emph{remove any overlapping target blocks}. 
   $\color{green}{\texttt{green}}$ patches are selected while $\color{gray}{\texttt{gray}}$ patches are removed. 
   }
   \label{fig:mask}
\end{figure}

\begin{figure}[t]
  \centering
   \includegraphics[width=0.95\linewidth]{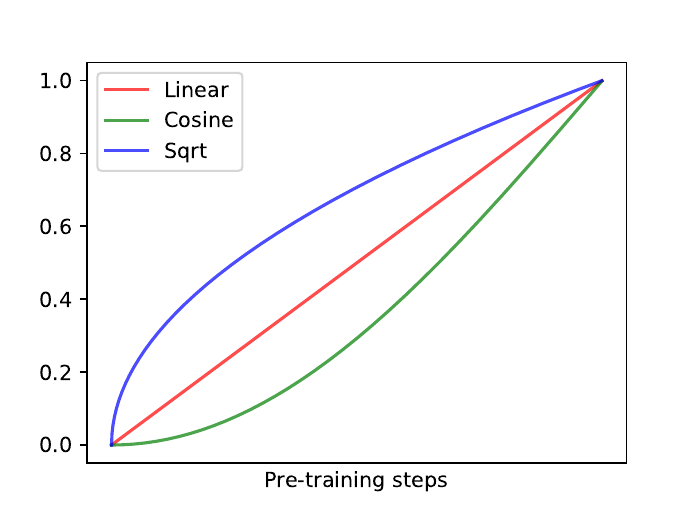}
   \vspace{-0.3cm}
   \caption{\textbf{Different curriculum function}. With the increase of the pre-training step, progressive function approx to hard (1) from easy (0) in different trends.}
   \label{fig:schedule}
\end{figure}

\paragraph{Curriculum masking strategy. }

Vanilla JEPA randomly masks out a block of spectrogram patches for the selection of context and target regions \cite{assran2023self}. 
As the spectrogram can be viewed as a 2D representation of audio along time and frequency with implicit entangles, it is reasonable to explore treating time and frequency differently during masking. 
Following \cite{huang2022masked},  we consider both random block masking without any prior and masking a portion of time and frequency of a spectrogram.  
It is believed that random block masking is comparably easier than time-frequency aware masking, as the model can guess the missing component by extrapolating nearby context, \emph{e.g.}, formants in vowels and frictional sounds in consonants around.  
That is, directly applying block masking will retain enough information in the time domain and frequency domain. It would be more difficult to completely obscure information from a certain paradigm. 
To this end, we carefully design two masking strategies with adjustable scale factors for specific masking ratios, as shown in Figure \ref{fig:mask}. 

Based on the above two masking strategies, we further propose an annealing dividing strategy with curriculum learning \cite{bengio2009curriculum,hacohen2019power,fei2021partially}. 
To be specific, we randomly decide whether to use random block or time-frequency aware masking at each training step and gradually anneal to the time-frequency aware masking method at the end of the training.
Figure \ref{fig:schedule} illustrate several popular progressive functions. 
Formally, given the image $I$ and the mask scale factor $r$,
we define block masking set $\mathcal{M}$ as follows: 
\begin{align} 
    p \sim  \text{Bernoulli}(f(s)) \\ 
\mathcal{M} =
\begin{cases}
\text{Time-frequency}(I, r) &\quad p= 1,\\
\text{Block}(I, r) &\quad p=0 \\
\end{cases}
\end{align} 
where $f(s)$ is the progressing function \cite{platanios2019competence}, and $s$ denotes the training step. We set $f(s) = \min (1, \sqrt{s \frac{1 - c_0^2}{S}} + c_0^2)$,
$c_0 > 0$ is set to 0.01 by default, as prior experiments shown a slight advantage compared with other schedule. 
Bernoulli($\cdot$) is the Bernoulli distribution with input
$f(s)$. 
The entire process can be seen in Algorithm \ref{algorithm}. 
Intuitively, a smaller value of $f(s)$ leads to easier reconstruction prediction. In this grade, $f(s)$ gradually increases from 0 to 1 automatically during pre-training, which results in a better representation performance.

\paragraph{Architecture.}
The employed architecture is reminiscent of the setting in \cite{huang2022masked,mae,assran2023self} and encompasses key components as: 
($\textbf{i}$) \emph{Context-encoder and target-encoder.} 
A-JEPA uses a stack of standard vision Transformers \cite{dosovitskiy2020image} as its encoder. 
The encoder exclusively processes non-masked patches. 
Target encoder, shares an identical structure and its weights are updated iteratively via an exponential moving average of the context encoder weights.
($\textbf{ii}$) \emph{Decoder} is also composed of standard Transformer blocks. The encoded patches from the encoder are padded with trainable masked tokens. After restoring the original time-frequency order in the audio spectrogram, we add fixed sinusoidal positional embeddings and feed the restored sequence into the decoder. At the topmost of the decoder, we add a linear head to predict and reconstruct the latent features.

\paragraph{Multi-mask objective.}
The objective is also simply to average the L2 distance between the output features of the predictor and the target encoder. 
To increase the efficiency of A-JEPA, we utilize a multi-masking strategy
\cite{caron2020unsupervised,baevski2023efficient}, which facilitates the amortization of the target computation expenses.

\paragraph{Fine-tuning with regularized masking. }
In the fine-tuning stage, we only keep and fine-tune the encoder and discard the decoder. An average pooling layer is applied followed by a linear layer on top in downstream tasks. 
On the other hand, certain works have attempted to integrate masking operations into the fine-tuning stage \cite{oneata2022improving,chen2020gridmask}. 
For example, SpecAug \cite{park2019specaugment} takes full-length input with the masked portion set to zero as data augmentation;  AudioMAE \cite{huang2022masked} selectively encounters a subset of real-valued input patches without the nullified ones with low masking ration. However, these methods may still lead to a discrepancy between actual conditions.

\begin{figure}[t]
  \centering
   \includegraphics[width=0.9\linewidth]{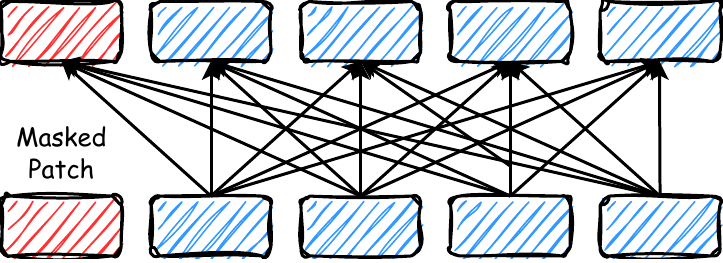}
   \caption{\textbf{Regularize patch masking during fine-tuning.} Masked patch is forbidden to attend to attention computation where its attention score is entirely contributed by others.  It manipulates the connections between patch tokens in self-attention via masking, where the networks are forced to exploit partial neighbors’ information to produce a meaningful representation.}
   \label{fig:rm} 
\end{figure}

In this study, we present regularized masking (RM), illustrated in Figure \ref{fig:rm}. 
In a nutshell, we modify the computing of self-attention to control the connections between different audio patch tokens, thereby influencing the attention scores and contextual semantic representation. 
Specifically, in the $l$-th layer, a certain percentage of audio patch tokens is first randomly selected. These tokens are excluded from the calculation of attention weights among the tokens, but entirely contributed by remaining others. 
In this manner, the networks are required to utilize partial neighbors’ attention information, and robust to acquire meaningful information.
We utilize RM in the fine-tuning phase while maintaining the attention calculation as the vanilla during testing.

\begin{algorithm}[t]
\caption{\small
{Curriculum Masking Strategy}}\label{algorithm}
\textbf{Input}: Image size ($w, h$), target scale $s_t$ and aspect ratio $r_t$, context scale $s_c$ and aspect ratio $r_c$, target mask number $N_t$, training step $s$, curriculum $f(\cdot)$; 
\begin{algorithmic}[1]
\STATE Context size $\mathcal{S}_c$ = Sample block Size($w, h, s_c, r_c$);
\STATE Target size $\mathcal{S}_t$ = Sample block Size($w, h, s_t, r_t$); 
\STATE $p \sim  \text{Bernoulli}(f(s))$;
\STATE Target mask list $M_t$, compliment target mask list $C_t$;
\FOR {$i$ in range($N_t$)}
    \IF{$p$ =1} 
        \STATE $m, c$ = Sample time-frequency mask($\mathcal{S}_t$);
    \ELSE
        \STATE $m, c$ = Sample block mask($\mathcal{S}_t$);
    \ENDIF 
    \STATE $M_t$ append $m$, $C_t$ append $c$;
\ENDFOR
\STATE Acceptable region $\mathcal{R} = C_t$;
\STATE Context mask $M_c$ = Sample context mask($\mathcal{S}_c$, $\mathcal{R}$);
\STATE \textbf{return} $M_t$ and $M_c$
    \end{algorithmic}
\end{algorithm}

\begin{table*}[t!]
\setlength\tabcolsep{3.0pt}

\begin{center}
\scalebox{0.99}{
\begin{tabular}{llllllllll} 
\multicolumn{1}{l}{\textbf{Model}} & \multicolumn{1}{l}{\textbf{Backbone}} & \textbf{PT-Data} & \textbf{AS-20K} & \textbf{AS-2M} & \textbf{ESC-50} & \textbf{SPC-2} & \textbf{SPC-1} & \textbf{SID} \\
\midrule
\multicolumn{3}{l}{\emph{No pre-training}} & \multicolumn{5}{l}{}\\
ERANN~\cite{verbitskiy2021eranns} & CNN & -& - & 45.0 & 89.2 & - & -& - \\
PANN~\cite{kong2019panns} & CNN & - & 27.8 & 43.1 & 83.3 & 61.8 &- &-  \\
\hline & \\[-2ex]
\multicolumn{3}{l}{\emph{In-domain self-supervised pre-training}}& \multicolumn{5}{l}{} \\
wav2vec 2.0~\cite{wav2vec2} & \footnotesize{Transformer} & LS & -  & - & - & - & 96.2\textsuperscript{*} & 75.2\textsuperscript{*}\\
HuBERT~\cite{Hsu2021HuBERTSS}  & \footnotesize{Transformer} & LS & - & - & - & - & 96.3\textsuperscript{*} & 81.4\textsuperscript{*} \\
Conformer~\cite{srivastava2021conformer} & \footnotesize{Conformer} & AS & - & 41.1 & 88.0 &  -  & -  & -\\
SS-AST~\cite{ssast} & ViT-B & AS+LS & 31.0 & - & 88.8 & 98.0 & 96.0 & 64.3 \\
\multicolumn{3}{l}{\textit{Mask and reconstruction works}}& \multicolumn{5}{l}{}\\ 
MaskSpec~\cite{chong2022masked} &ViT-B & AS & 32.3 & 47.1 & 89.6 & 97.7 & - & - \\
MAE-AST~\cite{baade2022} & ViT-B & AS+LS & 30.6 & - & 90.0 & 97.9 & 95.8 & 63.3 \\
{Audio-MAE} \cite{huang2022masked} & ViT-B & AS & {37.0} & {47.3} & {94.1}& {98.3}  & 96.9 & {94.8}\\
A-JEPA & ViT-B & AS &\textbf{38.4} &\textbf{48.6} &\textbf{96.3} &\textbf{98.5} &\textbf{97.7} &\textbf{95.8}\\
\hline & \\[-2ex]
\multicolumn{3}{l}{\emph{Out-of-domain supervised pre-training}}& \multicolumn{5}{l}{} \\
\color{gray}PSLA~\cite{gong2021psla} & \color{gray}EfficientNet & \color{gray}IN & \color{gray}31.9 & \color{gray}44.4 & - & \color{gray}96.3 & - & \color{gray}- \\
\color{gray}AST~\cite{gong2021ast} & \color{gray}DeiT-B & \color{gray}IN &   \color{gray}34.7 & \color{gray}45.9 & \color{gray}88.7 & \color{gray}98.1 & \color{gray}95.5 & \color{gray}41.1 \\ 
\color{gray}MBT~\cite{Nagrani21c} & \color{gray}ViT-B & \color{gray}IN-21K &  \color{gray}31.3 & \color{gray}44.3 & - & - & - & - \\
\color{gray}HTS-AT~\cite{chen2022hts} & \color{gray}Swin-B & \color{gray}IN & - & \color{gray}47.1 & \color{gray}97.0& \color{gray}98.0 & - & - \\
\color{gray}PaSST~\cite{paast} & \color{gray}DeiT-B & \color{gray}IN & - & \color{gray}47.1 & \color{gray}96.8 & - & - & - \\
\end{tabular}}
\caption{
\textbf{Comparison with other state-of-the-art audio representation models} on audio and speech classification tasks. 
We use the mAP (\%) metric for AS and the accuracy metric for ESC/SPC/SID tasks. 
For pre-training (PT) dataset, we simplify as AudioSet (AS), LibriSpeech (LS), and ImageNet (IN). 
\textsuperscript{$\dagger$} denotes Fine-tuning results with additional supervised training on AS-2M. 
We {\color{gray}gray-out} models pre-trained with external non-audio datasets, e.g., ImageNet.
The best single models in AS-2M are compared in no ensembles. \textsuperscript{*} represents linear evaluation results from corresponding papers. 
}
\label{tab:sota}
\end{center}
\end{table*}

\section{Experiments}

We perform an extensive evaluation on six tasks, including audio classification on AudioSet (AS-2M, AS-20K), Environmental Sound Classification (ESC-50), and speech classification on Speech Commands (SPC-1 and SPC-2) and VoxCeleb (SID). We use AudioSet for ablation studies. The implementation details can be seen in Appendix A. 

\subsection{Experimental settings.}

For a comparable comparison, similar to \cite{huang2022masked}, we utilize the following tasks with datasets: 

\noindent
(\textbf{i}) \texttt{AudioSet}~\cite{gemmeke2017audio} (AS-2M, AS-20K)
comprises approximately 2 million 10 seconds that are utilized for audio classification purposes. Each clip in the dataset is weakly annotated for 527 types of audio events ~\cite{tagging_right,vggish,hershey2021benefit}, with the possibility of multiple events occurring within a single clip. The dataset consists of a full training set, which is further divided into two subsets: a class-wise balanced set containing 22,176 clips, and an unbalanced set containing 2,042,985 clips. Additionally, an evaluation set with 20,383 clips is provided for testing purposes. To conduct our experiments, we obtained and processed a subset of the dataset, comprising 1.96M clips from the unbalanced training set, 21k clips from the balanced training set, and 19k clips from the evaluation set. 
We use the union of unbalanced and balanced training audio for pre-training and fine-tuning for AS-2M; We use AS-2M for pre-training and the 20K balanced set for fine-tuning in the AS-20K experiment.
We report the testing mAP on the 19K eval set following AST~\cite{gong2021ast}. 

\noindent
(\textbf{ii}) \texttt{Environmental Sound Classification} 
(ESC-50)~\cite{piczak2015dataset} is a collection of 2,000 environmental sound recordings for audio classification. Each lasts 5 seconds and there are 50 classes in ESC. We report accuracy under 5-fold cross-validation with the same split used by~\cite{gong2021ast}. 

\noindent
(\textbf{iii}) \texttt{Speech Commands} (SPC-2, SPC-1)~\cite{speechcommandsv2} pertain two keyword spotting tasks.
SPC-2 encompasses a set of 35 speech commands. 
The training set is 84,843 and the testing and validation sets are with sizes 11,005, and 9,981 respectively. Each with a duration of 1 second. 
SPC-1 encompasses 10 classes of keywords, 1 silence class, and 1 unknown class that includes all remaining 20 common speech commands. We employ the data and partitioning from SUPERB~\cite{yang21c_interspeech} benchmark to report the testing accuracy.

\noindent
(\textbf{iv}) \texttt{VoxCeleb} (SID)~\cite{Nagrani2020VoxcelebLS} comprises 150K utterances attributed to 1,251 distinct speakers. 
The speaker identification task (SID) involves to classify the utterances to identify their original speaker. 
We use the V1 version for training, validation, and testing sets with 128k, 6k, and 8k, and report the testing accuracy metric.

\begin{table}[t!]
\centering
\begin{tabular}{llcc}

                    \textbf{Backbone} & \textbf{\#Params} & \textbf{AS-20K} & \textbf{AS-2M}  \\
                \midrule
               ViT-S &  22M & 33.7 & 46.1  \\ 
                 ViT-B & 86M &38.4 &48.6 \\
                 ViT-L & 304M & \textbf{38.8} & \textbf{48.8}  \\
                 
            \end{tabular}
    \caption{\textbf{Ablating model size.} Evaluating the impact of model size on downstream tasks. A-JEPA benefits from larger models.}
    \label{tab:ablation:model_size}
\end{table}

\subsection{Comparison with the State-of-the-art}

Table~\ref{tab:sota} presents a comprehensive comparison of A-JEPA against previous state-of-the-art models for audio representation learning. The analysis is divided into three distinct groups. 
To ensure a fair evaluation, our primary focus lies on models within the middle group, which have undergone self-supervised pre-training using in-domain audio-only datasets
In addition, we include other models without any pre-training at the top group and models with supervised pre-training on out-of-domain ImageNet at the bottom group, where the latter comprises the previous leading systems on the respective datasets. 
Among the models with in-domain self-supervised pre-training, A-JEPA, pre-trained on AudioSet, exhibits the highest performance across all tasks. Notably, its mAP score of 38.4 on the AudioSet-20K dataset surpasses all alternative approaches, including previous masking-reconstruction works Conformer~\cite{srivastava2021conformer}, SS-AST~\cite{ssast} and AudioMAE~\cite{huang2022masked}. 
In the lowermost group of Table~\ref{tab:sota}, A-JEPA also demonstrates superior performance compared to previous state-of-the-art models that employed ImageNet supervised pre-training. 
It is worth noting that A-JEPA, the proposed approach, does not rely on external data or labels from unrelated domains.
More encouragingly, as indicated in the experiments conducted in~\cite{kong2019panns}, there is a potential for further mAP enhancement for A-JEPA if audio data with a sampling rate of 32K becomes available. 
The advantage is still maintained for the speech tasks, including SPC-1, SPC-2, and SID. 
In summary, A-JEPA, which leverages audio-only pre-training from scratch using AudioSet, achieves commendable performance in both audio and speech classification tasks through the incorporation of feature alignment.

\begin{figure*}[t]
  \centering
   \includegraphics[width=1.\linewidth]{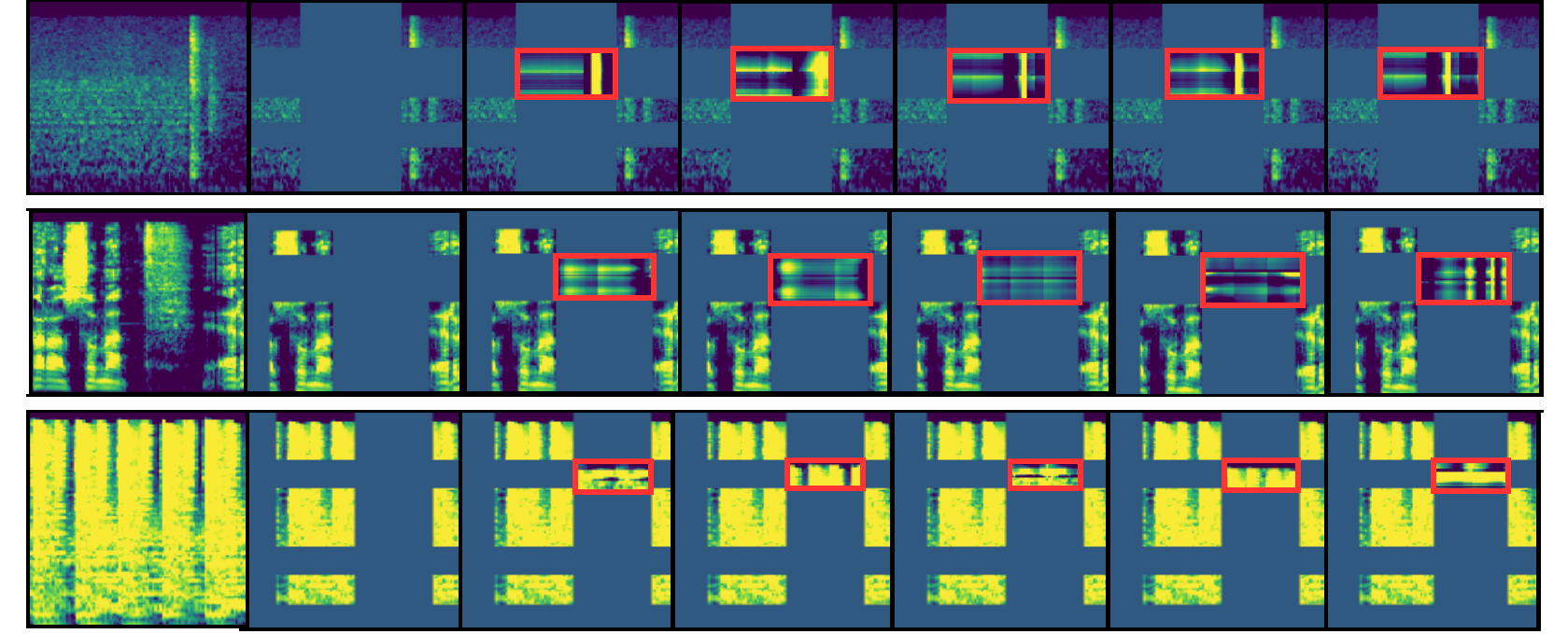}
   \caption{\textbf{Visualization of A-JEPA predictor representations.} The first column presents the original audio spectrogram, while the second column displays the context audio spectrogram, which is processed with a pre-trained A-JEPA ViT-B encoder. 
   $\color{red}{\texttt{red}}$ bounding boxes, in subsequent columns, showcase samples created from a generative model. It decoded the output of the pre-trained A-JEPA predictor, conditioned on positional mask tokens corresponding to the location of the bounding box. It is worth noting that qualities shared among these samples indicate the information contained in the A-JEPA prediction.
   }
   \label{fig:data_case}
\end{figure*}

\subsection{Model Analysis}

\paragraph{Model scality.}
We study the various design choices pertaining to encoder architectures in the A-JEPA.
Table~\ref{tab:ablation:model_size} illustrates the trade-off between encoder model size and its performance. 
It is observed that larger encoder models tend to exhibit superior performance, albeit at the expense of increased computational requirements and memory utilization. 
Moreover, it is noted that the accuracy improvement of ViT-L over ViT-B/S is particularly pronounced when applied to the smaller and more balanced AS-20K dataset. Additionally, the performance disparity between ViT-S and ViT-B can be considerably diminished, e.g., from 4.7 to 2.5 mAP, by employing fine-tuning techniques with a greater volume of in-domain data, e.g., from AS-20K to AS-2M, as observed in \cite{huang2022masked}.

\begin{table}[t!]
\centering
    \begin{tabular}{lccc}
    
        \textbf{Mask} & \textbf{Backbone} & \textbf{Mask Ratio} & \textbf{mAP}\\
        \midrule
        \texttt{Curriculum} &ViT-B &0.75 & \textbf{48.6} \\
        \texttt{Random} &ViT-B & 0.75 & 47.8 \\
        \texttt{Inverse} &ViT-B  & 0.75 & 46.3 \\
         
    \end{tabular}   
    \caption{\textbf{Ablating masking strategy in pre-training}. The proposed \texttt{curriculum} masking strategy is preferred.}
    \label{fig:masking_comparison}
\end{table}      

\begin{table}[t!]
\centering
    \begin{tabular}{lccc}
    
        \textbf{Model} &\textbf{Backbone} & \textbf{Masking Ratio} & \textbf{mAP}\\
        \midrule
        A-JEPA &ViT-B &0.1 & \textbf{48.6} \\
       \emph{w/o} RM &ViT-B & 0.1 & 48.1 \\
         
    \end{tabular}   
    \caption{\textbf{Ablating regularized masking (RM) in fine-tuning}. The proposed regularized masking strategy in fine-tuning is helpful for guiding A-JEPA to learn more robust audio representations.}
    \label{tab:ft}
\end{table}

\paragraph{Masking strategies in pre-training and fine-tuning.}
We also present a comparison of different pre-training masking strategies for A-JEPA in In Table ~\ref{fig:masking_comparison}.
The results obtained from our experiments reveal that curriculum masking outperforms random block masking. This finding suggests that a guided approach to time-frequency aware masking, yields superior performance. 
Moreover, inverse strategy denotes that first time and frequency and then random maksing, which results in a more significant performance drop, demonstrate the effectiveness of our method again. 
In general, we observe that for task-agnostic pre-training, curriculum masking from easy to hard with a high ratio is preferred. 
On the other hand, when it comes to fine-tuning, as illustrated in Table~\ref{tab:ft}, employing regularized patch masking with lower ratios achieves better performance in downstream tasks.

\paragraph{Predictor depth and width.} 
The impact of decoder depth on mean average precision (mAP) is evaluated in Table \ref{depth}. A 16-layer decoder, being deeper than its shallower counterparts, exhibits superior performance. 
Furthermore, Table \ref{width} presents a comparison of decoder widths, specifically the embedding dimension. The results indicate that a 512-dimension decoder strikes a favorable balance between computational requirements and performance, as increasing the width beyond this threshold does not yield significant improvements.

\begin{table}[t!]
\centering
    \begin{tabular}{lccc}
            \textbf{Depth} &\textbf{Backbone} &\textbf{Epoch} & \textbf{mAP}\\
        \midrule
        8 &ViT-B &24 & 48.3 \\
        16 &ViT-B&24 &\textbf{48.6} \\
          
    \end{tabular}   
    \caption{\textbf{Ablating the predictor depth.} Increasing the predictor depth leads to a prominent performance improvement of the pre-trained audio representations.}
    \label{depth}
\end{table}

\begin{table}[t!]
\centering
    \begin{tabular}{lccc}
        \textbf{Width} &\textbf{Backbone} &\textbf{Epoch} & \textbf{mAP}\\
        \midrule
        256&ViT-B&24 & 48.0 \\
        512&ViT-B&24 &\textbf{48.6}\\
        768&ViT-B&24 &48.5\\
          
    \end{tabular}   
    \caption{\textbf{Ablating the predictor width.} Having a width bottleneck in the predictor improves the downstream performances.}
    \label{width}
\end{table}

\begin{figure}[t]
  \centering
   \includegraphics[width=0.95\linewidth]{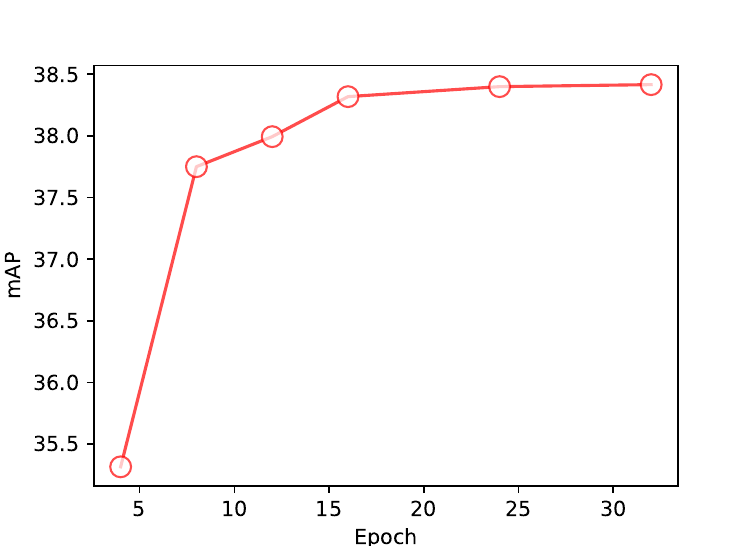}
   \caption{\textbf{Ablating pre-training epochs on ViT-B backbone in AudioSet.} Training for longer is beneficial for downstream task performance.}
   \label{fig:epoch}
\end{figure}

\paragraph{Pre-training data size and epochs.}
Figure \ref{fig:data_case} depicted the influence of pre-training dataset size. A subset of data is randomly sampled from originals for pre-training.   
It is observed that the performance of the model consistently improves with an increase in the amount of data used for pre-training.
Meanwhile, as shown in Figure \ref{fig:epoch}, we also plot the mAP score at different training epochs. 
It is evident that prolonging the training duration yields favorable outcomes; however, the performance reaches a plateau after the 24th epoch.

\begin{figure}[t]
  \centering
   \includegraphics[width=0.95\linewidth]{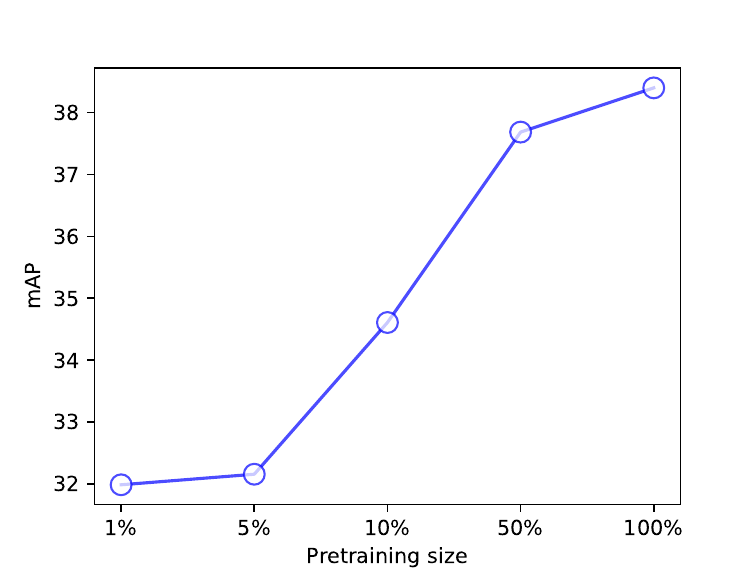}
   \caption{\textbf{Ablating pre-training data size.} When increasing the size of the pre-training dataset, we see a significant performance improvement in audio representation for the ViT-B model. 
   }
   \label{fig:size}
\end{figure}

\subsection{Predictor Visualizations}
The purpose of the predictor component in I-JEPA is to utilize the output of the context-encoder and provide predictions for the representations of a target object at a specific location, as specified by the positional mask tokens. In this section, we aim to analyze whether the predictor conditioned on the positional mask tokens is learning to correctly capture positional uncertainty in the target in the audio spectrogram. 
Specifically, we qualitatively visualize the outputs of the predictor referring to \cite{assran2023self}.  After pretraining, we freeze the context-encoder and predictor weights, and train a decoder following the RCDM framework \cite{bordes2021high} to map the average pool of the predictor outputs back to pixel space. The outputs of the decoder for different random seeds are illustrated in Figure \ref{fig:data_case}. 
Shared characteristics observed across multiple samples indicate information that is conveyed within the average-pooled predictor representation. It is noteworthy that the A-JEPA predictor successfully captures positional uncertainty and generates high-level audio components with an accurate pose.

\section{Conclusion}

We have conducted an investigation into an extension of JEPA to audio data. Our A-JEPA is designed to reconstruct masked spectrogram patches from audio recordings within a latent space and achieves superior performance on multiple audio and speech classification tasks. 
We have made three noteworthy observations: First, a straightforward application of JEPA yields remarkable outcomes for audio spectrograms. Second, we find that it is possible to enhance the quality of learned representations with a time-frequency aware masking strategy progressing from easy to hard. 
Third, we show that regularized masking, instead of directly dropping or setting to zero, can be applied to fine-tuning, contributing to accuracy improvement.  
In the future, we intend to explore multi-modal self-supervised learning with a joint latent embedding, \emph{e.g.}, video and text modals, to provide a better formulation of audio representation guidance. 

\bibliographystyle{named}
\bibliography{ijcai23}

\clearpage
\appendix

\section{Implementation Details}

For model structure, we use a vanilla 12-layer ViT-B by default as the Transformer encoder and use a 16-layer vanilla Transformer as the decoder. The design of other ViT model sizes is identical to \cite{huang2022masked}.
Following~\cite{gong2021ast,Nagrani21c}, we convert raw waveform, pre-processed as mono channel under 16,000 sampling rate, into 128 Kaldi-compatible Mel-frequency bands ~\cite{povey2011kaldi} with a 25ms Hanning window that shifts every 10 ms. 
In the case of a 10-second recording in AudioSet, the resulting spectrogram is of 1$\times$1024$\times$128 dimension.
For patch embedding, we employ convolutional kernels with a size of $(16,16)$ and apply a stride in both time and frequency, ensuring that the resulting patches are non-overlapping. We sample 4 possibly overlapping target random blocks with a scale in the range (0.15, 0.2) and aspect ratio in the range (0.75, 1.5); sample time-frequency aware 3 target blocks with scale in the range (0.05, 0.075). We sample 1 context block mask with a random scale in the range (0.85, 1.0) and unit aspect ratio. We subsequently eliminate specific regions, different for random and time-frequency, in the context block mask that overlaps with any of the target block masks. 
These hyperparameters are tested in prior experiments to balance a patch available number. 
The context-block mask and target-block masks are sampled independently for each image in the mini-batch.


The pre-training phase involves the utilization of AudioSet-2M, wherein we conduct a random iteration over all audio recordings.
We train for 24 epochs with a batch size of 512, coupled with a learning rate of 2e-4.  
For each audio, we randomly sample the starting time, cyclically extract a 10-second audio segment, and randomly jitter its magnitude by up to $\pm$ 6dB. 
During fine-tuning, we employ a 10\% regularized patch masking ratio by default.
For the supervised fine-tuning on AudioSet-2M, to address the issue of imbalanced training sample sizes across classes, we follow the common practice of using a weighted sampling to balance the classes during training \cite{huang2022masked}.  
In each epoch, we sample 200K instances, approximately 10\% of AudioSet-2M, without replacement. We fine-tune for 100 epochs, which aggregate to 10 full epochs of AudioSet-2M. 
This sampling procedure ensures that the probability of selecting an instance is inversely proportional to the occurrences of its corresponding class within the dataset. 
For the smaller balanced AudioSet-20K, we fine-tune for 60 epochs without weighted sampling. We use the Sqrt progressive function based on prior observations.

\section{Pytorch Code for Time-frequency Aware Masking}
We provide the pytorch implementation for time-frequency aware masking.
\begin{listing}[H]%
\caption{Pytorch code for time-frequency aware masking}%
\label{lst:listing}%
\begin{lstlisting}[language=Haskell]
def sampe_time_frequency_mask(b_size, acceptable_regions=None, min_ratio=0.35):
    """b_size: sampled block size."""
    h, w = b_size
    def constrain_mask(mask, tries=0):
        """ Helper to restrict given mask to acceptable regions """
        N = max(int(len(acceptable_regions)-tries), 0)
        for k in range(N):
            mask *= acceptable_regions[k]

    valid_mask = False
    while not valid_mask:
        # Sample block top-left corner
        top = torch.randint(0, height - h, (1,))
        left = torch.randint(0, width - w, (1,))
        mask = torch.zeros((height, width), dtype=torch.int32)
        mask[top:top+h, left:left+w] = 1
        # Constrain mask to a set of acceptable regions
        if acceptable_regions is not None:
            constrain_mask(mask, tries)
        mask = torch.nonzero(mask.flatten())
        # If mask too small try again
        valid_mask = len(mask) > min_ratio * h * w

    mask_complement = torch.ones((height, width), dtype=torch.int32)
    mask_complement[top:top+h, left:left+w] = 0

    # Remove the time and frequency part
    mask_complement_tf = torch.ones((height, width), dtype=torch.int32)
    mask_complement_tf[top:top+h, :] = 0
    mask_complement_tf[:, left:left+w] = 0
    return mask, mask_complement, mask_complement_tf
\end{lstlisting}
\end{listing}

\end{document}